\documentclass[aps,prl,twocolumn,showpacs,superscriptaddress]{revtex4-2}

\usepackage{graphicx}
\usepackage{amssymb}
\usepackage{amsmath}
\usepackage{color,xcolor}
\usepackage{bm}
\usepackage{physics}
\usepackage{ulem}

\usepackage{changes}

\begin{document}


\title{Mobile Exceptional Points Generate Momentum-Space Switching Domains}

\author{Jung-Wan Ryu}
    \address{Center for Trapped Ion Quantum Science, Institute for Basic Science (IBS), Daejeon 34126, Republic of Korea}
    \address{IBS school of Basic Science, Korea University of Science and Technology (UST), Daejeon 34113, Republic of Korea}

\author{Chang-Hwan Yi}
\email{yichanghwan@hanmail.net}
    \address{Center for Theoretical Physics of Complex Systems, Institute for Basic Science (IBS), Daejeon 34126, Republic of Korea}
    \address{Department of Physics, Pukyong National University, Busan 48513, Republic of Korea}

    \address{Department of Physics, Hanyang University, Seoul 04763, Republic of Korea}

    
\date{March 9, 2026 - \today}

\begin{abstract}
Exceptional points (EPs), non-Hermitian degeneracies where both eigenvalues and eigenvectors coalesce, play a central role in the topology of non-Hermitian spectra. Recent advances have enabled the controlled creation and manipulation of EPs in a wide range of physical systems, raising the question of what new band topology emerges when EPs become mobile under cyclic modulation. Here we show that mobile EPs generate momentum-space switching domains that partition the Brillouin zone into regions with distinct band-switching behavior. Using a minimal two-band lattice model, we introduce a band-permutation invariant that determines whether eigenmodes exchange after one modulation cycle. The boundaries between switching regions arise from the projection of EP trajectories in an extended parameter space combining crystal momentum and the modulation parameter. As the modulation strength increases, the switching domains expand and eventually cover the entire Brillouin zone, resulting in global band switching. The predicted switching-domain structure is further demonstrated in a photonic crystal with lossy materials. These results open a new avenue within non-Hermitian topology by enabling the engineering of EP-driven phenomena through their controlled motion.
\end{abstract}

\maketitle

\textcolor{blue}{\textit{Introduction -}}
Non-Hermitian systems exhibit spectral degeneracies known as exceptional points (EPs), where both eigenvalues and eigenvectors coalesce and the complex energy spectrum acquires a multi-sheeted topology \cite{Kato1976perturbation, Heiss1990avoided, Heiss_2004exceptional, Ryu2009coupled, Dietz2011exceptional, Xu2016topological, Miao2016orbital, Ding2016emergence, Chen2017exceptional}. These singular degeneracies give rise to unusual wave phenomena and rich spectral topology that have attracted extensive attention \cite{Heiss1999phases, Gao2015Observation, Zhen2015spawning, Shen2018topological, Luitz2019exceptional, Chen2020revealing, Tang2020exceptional, Bergholtz2021exceptional, Yang2021fermion, Wang2021topological, Schumer2022topological, Zhang2023observation, Arkhipov2024Restoring, Chen2024machine, Erb2024novel, Yi2025robust}. Recent experimental advances have enabled precise control of non-Hermitian systems through mechanisms such as engineered gain and loss, asymmetric coupling, and temporal modulation \cite{Guo2009observation, Rueter2010observation, El-Ganainy2018non, Park2022revealing, Ergoktas2022topological, Shen2023nonreciprocal, Zhao2024exceptional}. As a result, EPs can now be directly observed and systematically engineered across a wide range of platforms, including photonic structures, microwave resonators, mechanical metamaterials, and electronic circuits \cite{Schindler2011experimental, Peng2014parity, Shi2016accessing, Brandenbourger2019nonreciprocal, Helbig2020generalized}. These advances further enable EPs to be created, moved, merged, and annihilated through controlled parameter tuning \cite{Zhang2019dynamically, Miri2019exceptional, Oezdemir2019parity, Liu2020efficient, Gu2021controlling, Yu2021general}.

A hallmark consequence of EP physics is eigenmode switching under cyclic parameter evolution \cite{Dembowski2001experimental, Dembowski2004encircling, Ryu2012analysis, Lee2012geometric, Doppler2016dynamically, Zhong2018winding, Yoon2018time, Li2020hamiltonian}. When system parameters trace a closed loop that encircles an EP, the eigenvalues follow the multi-sheeted topology of the associated Riemann surface, leading to the exchange of eigenmodes after one cycle. This behavior reflects the topological nature of EPs as codimension-two degeneracies in complex spectra, which act as robust defects that can only be created or annihilated in pairs. While eigenmode switching around fixed EPs is by now well understood, most previous studies have focused on closed loops in parameter space, where the central question is whether a given loop encloses EPs.

Here we investigate a distinct regime in which EPs themselves become mobile under periodic modulation of a control parameter. In this regime, eigenmode switching is no longer determined solely by the external parameter loop but instead becomes momentum dependent across the Brillouin zone. We show that the motion of EPs reorganizes the band topology by generating momentum-space domains that partition the Brillouin zone into regions with distinct switching behavior. The central question is thereby reformulated: instead of asking whether a loop encloses an EP, one asks for which crystal momenta the spectrum switches branches during a modulation cycle. The resulting switching pattern defines domains in momentum space whose boundaries appear as closed curves in the Brillouin zone and correspond to the projection of EP trajectories in an extended parameter space that includes both crystal momentum and the cyclic control parameter.

\begin{figure}
\centering
\includegraphics[width=0.6\linewidth]{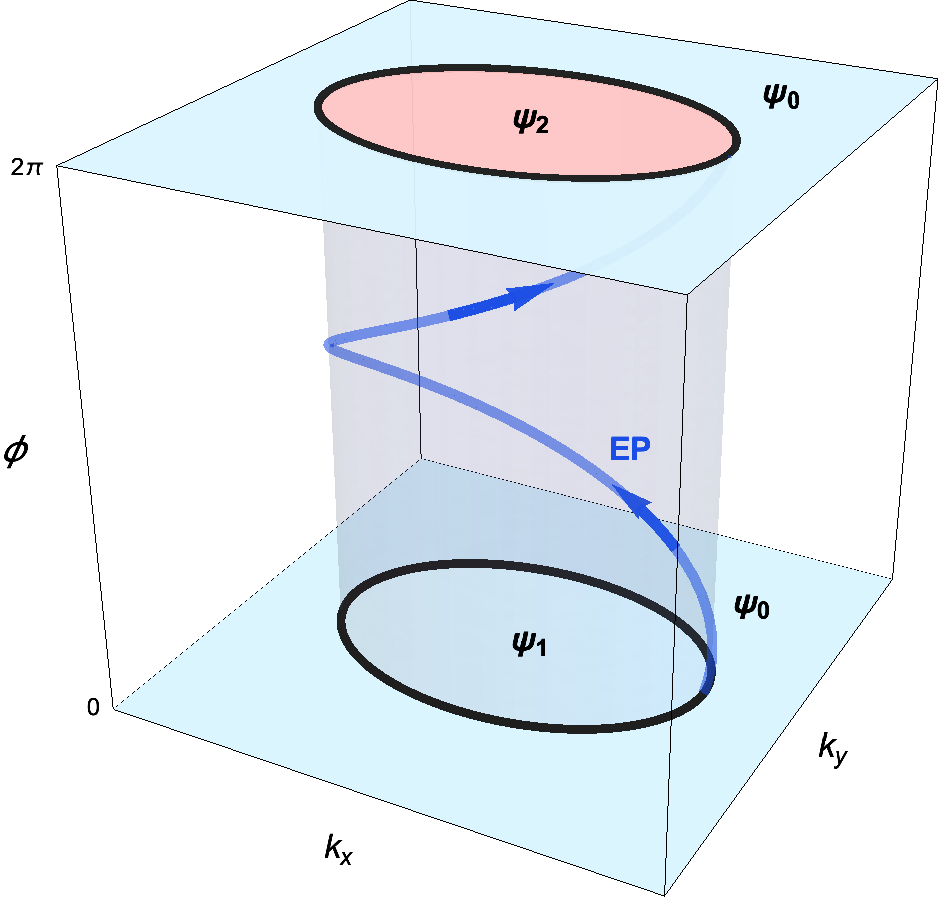}
\caption{
Schematic illustration of momentum-space switching domains generated by mobile EPs. 
An EP traces a trajectory in the extended parameter space $(k_x,k_y,\phi)$ as the control parameter $\phi$ evolves during one cycle. 
The projection of this trajectory onto the $(k_x,k_y)$ plane forms a closed loop that separates momenta with different dynamical behavior. 
Crystal momenta inside the loop undergo eigenmode switching ($\psi_1\!\rightarrow\!\psi_2$) after one cycle, whereas those outside return to the same state ($\psi_0\!\rightarrow\!\psi_0$). 
This geometric picture illustrates how the motion of EPs organizes momentum-space switching domains.
}
\label{fig:schematic}
\end{figure}

The underlying mechanism is illustrated schematically in Fig.~\ref{fig:schematic}. As the control parameter evolves over one cycle, an EP traces a trajectory in the extended parameter space $(k_x,k_y,\phi)$. The projection of this trajectory onto the Brillouin zone separates momenta that undergo eigenmode switching from those that return to their original state. This geometric picture provides a natural organizing principle for non-Hermitian band topology, in which the motion of EPs governs momentum-space switching domains and determines eigenmode permutations across the Brillouin zone. We further demonstrate that this mechanism extends beyond minimal models and can be realized in experimentally relevant systems.

\begin{figure*}
\centering
\includegraphics[width=1.0\linewidth]{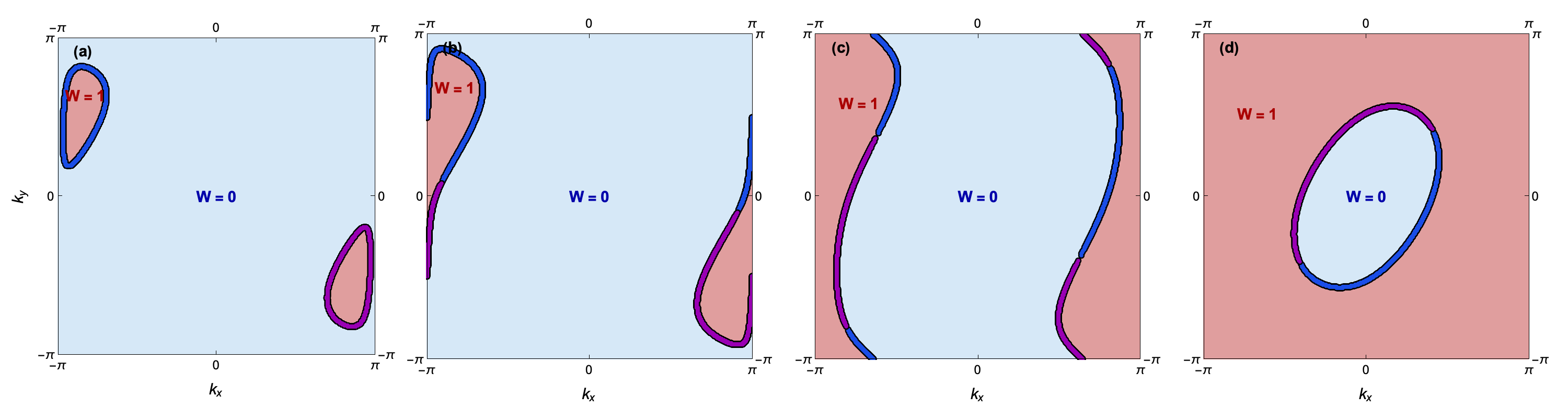}
\caption{Momentum-space switching domains generated by mobile EPs. 
The Brillouin zone is partitioned into regions characterized by the band-permutation invariant $W(\mathbf{k})$, where $W=1$ indicates eigenmode exchange after one modulation cycle and $W=0$ indicates no switching. 
The colored background shows the distribution of $W(\mathbf{k})$ in the Brillouin zone. 
Blue and purple curves denote the projected trajectories of the two EPs as the control parameter $\phi$ evolves over one cycle. 
(a) For small modulation amplitude $\mu$, switching occurs only in isolated regions of the Brillouin zone. 
(b) As $\mu$ increases, these regions expand and merge. 
(c) With further increase of $\mu$, the switching domains evolve to wrap around the Brillouin-zone torus. 
(d) At larger $\mu$, the switching region covers almost the entire Brillouin zone. 
The switching-domain boundaries coincide with the projections of the EP trajectories.}
\label{fig:domains}
\end{figure*}

\textcolor{blue}{\textit{Model -}}
We consider a minimal two-band non-Hermitian lattice model depending on the crystal momentum $\mathbf{k}=(k_x,k_y)$ and a cyclic control parameter $\phi$,
\begin{equation}
H(\phi,\mathbf{k})=
\begin{pmatrix}
\epsilon_0 & t_x+t_y \\
t_x e^{ik_x}+t_y e^{ik_y}+\mu e^{i\phi} & -\epsilon_0
\end{pmatrix}.
\end{equation}
Here $\epsilon_0$ denotes the onsite potential difference between the two sublattices, while $t_x$ and $t_y$ represent the hopping amplitudes along the $x$ and $y$ directions. The term $\mu e^{i\phi}$ introduces a phase-modulated coupling with amplitude $\mu$, and $\phi\in[0,2\pi]$ serves as the cyclic control parameter. This minimal model captures the essential features of non-Hermitian band structures with mobile EPs and provides a generic framework for understanding the resulting switching-domain physics.

The eigenvalues take the form $E_\pm(\mathbf{k},\phi)=\pm\Delta(\mathbf{k},\phi)$, where
\begin{equation}
\Delta(\mathbf{k},\phi)=
\sqrt{\epsilon_0^2+(t_x+t_y)\bigl(t_x e^{ik_x}+t_y e^{ik_y}+\mu e^{i\phi}\bigr)} .
\end{equation}
For a fixed momentum $\mathbf{k}$, varying $\phi$ from $0$ to $2\pi$ generates a closed trajectory of $\Delta(\mathbf{k},\phi)$ in the complex plane. Eigenmode exchange occurs when this trajectory winds around $\Delta=0$, which corresponds to an EP where both eigenvalues and eigenvectors coalesce. Thus, winding of $\Delta(\mathbf{k},\phi)$ around the origin indicates that the parameter cycle encloses an EP in the extended parameter space.

To characterize this behavior, we define a band-permutation invariant
\begin{equation}
W(\mathbf{k})=\frac{1}{2\pi i}\oint_0^{2\pi}
\partial_\phi \log \left[\Delta^2(\mathbf{k},\phi)\right]\, d\phi ,
\end{equation}
which measures the winding number of $\Delta^2(\mathbf{k},\phi)$ around the origin in the complex plane during one modulation cycle. 
Here, the squared quantity $\Delta^2(\mathbf{k},\phi)$ is used to ensure single-valuedness over the cycle. 
Since $E_\pm=\pm\Delta$, a winding number $W(\mathbf{k})=1$ implies that $\Delta^2(\mathbf{k},\phi)$ encircles the origin, 
which corresponds to a sign change of $\Delta(\mathbf{k},\phi)$ after one cycle and leads to eigenvalue permutation. 
In the present two-band system the invariant takes the values $W(\mathbf{k})=0$ or $1$, indicating the absence or presence of eigenmode switching.

The winding number can change only when the spectral trajectory crosses $\Delta=0$, where the spectral gap closes and an EP appears in the extended parameter space $(k_x,k_y,\phi)$. As the control parameter varies, the EPs trace trajectories in this three-dimensional space. The projection of these trajectories onto the Brillouin zone forms closed curves that separate regions with different values of $W(\mathbf{k})$. Eliminating $\phi$ from the EP condition yields the projected locus of EP trajectories in the Brillouin zone [SM-A],
\begin{equation}
\bigl|
\epsilon_0^2+(t_x+t_y)(t_x e^{ik_x}+t_y e^{ik_y})
\bigr|
=|t_x+t_y|\mu .
\end{equation}
These projected trajectories thus coincide with the momentum-space domain boundaries.

\textcolor{blue}{\textit{Momentum-space switching domains -}}
We compute the band-permutation invariant $W(\mathbf{k})$ to determine how eigenmode switching depends on crystal momentum [SM-B]. The resulting map, shown in Fig.~\ref{fig:domains} for several values of the modulation strength $\mu$, reveals momentum-space domains characterized by distinct values of $W(\mathbf{k})$.

Figure~\ref{fig:domains} further shows that these domains originate from the motion of EPs. Regions with $W(\mathbf{k})=1$ correspond to spectral trajectories $\Delta(\mathbf{k},\phi)$ that wind around the origin during one modulation cycle, resulting in eigenmode exchange, whereas regions with $W(\mathbf{k})=0$ exhibit no switching. The domain boundaries form closed curves in momentum space and coincide with the projections of EP trajectories obtained from the EP condition. Across these curves, $W(\mathbf{k})$ changes discontinuously, marking the transition between switching and nonswitching behavior. This domain structure thus reflects the motion of EPs in the extended parameter space $(k_x,k_y,\phi)$ [SM-C].

The switching domains evolve systematically as the modulation strength $\mu$ increases. For small $\mu$, switching occurs only in isolated regions of the Brillouin zone [Fig.~\ref{fig:domains}(a)]. As $\mu$ increases, these regions expand and merge [Fig.~\ref{fig:domains}(b)]. With further increase of $\mu$, the switching domains wrap around the Brillouin-zone torus [Fig.~\ref{fig:domains}(c)], and eventually cover almost the entire Brillouin zone [Fig.~\ref{fig:domains}(d)]. The above results can be summarized as follows. For a single EP forming a simple closed loop, the interior region corresponds to eigenmode switching, while the exterior does not. When multiple EPs collectively form a single loop, however, the boundary still separates distinct domains, but the assignment of switching behavior to each side is not determined by a simple inside-outside relation.

For sufficiently large $\mu$, the remaining $W(\mathbf{k})=0$ region disappears and the invariant becomes $W(\mathbf{k})=1$ throughout the Brillouin zone, resulting in global band switching for all crystal momenta. In this regime, the EPs no longer exist in the extended parameter space $(k_x,k_y,\phi)$. Nevertheless, the switching behavior can still be understood from the global topology of the complex spectrum. Since the eigenvalues are given by $E_\pm=\pm\Delta(\mathbf{k},\phi)$, eigenmode switching is determined by whether the closed trajectory $\Delta(\mathbf{k},\phi)$ winds around the origin during one modulation cycle. Equivalently, this can be described in terms of the parity of intersections with lines associated with eigenmode exchange: an odd number of such intersections leads to switching ($W(\mathbf{k})=1$), whereas an even number yields no net exchange ($W(\mathbf{k})=0$). Related structures in the complex spectrum were analyzed in multiband non-Hermitian systems \cite{Ryu2025pseudo}. The dynamical evolution of this spectral structure during the modulation cycle is illustrated in the supplementary animations. A complementary analysis of the EP regimes of the model, including the parameter window where EPs can exist, is presented in the Supplementary Material (see Supplementary Figs.~S2 and S3) [SM-D].

\begin{figure}
\centering
\includegraphics[width=0.95\linewidth]{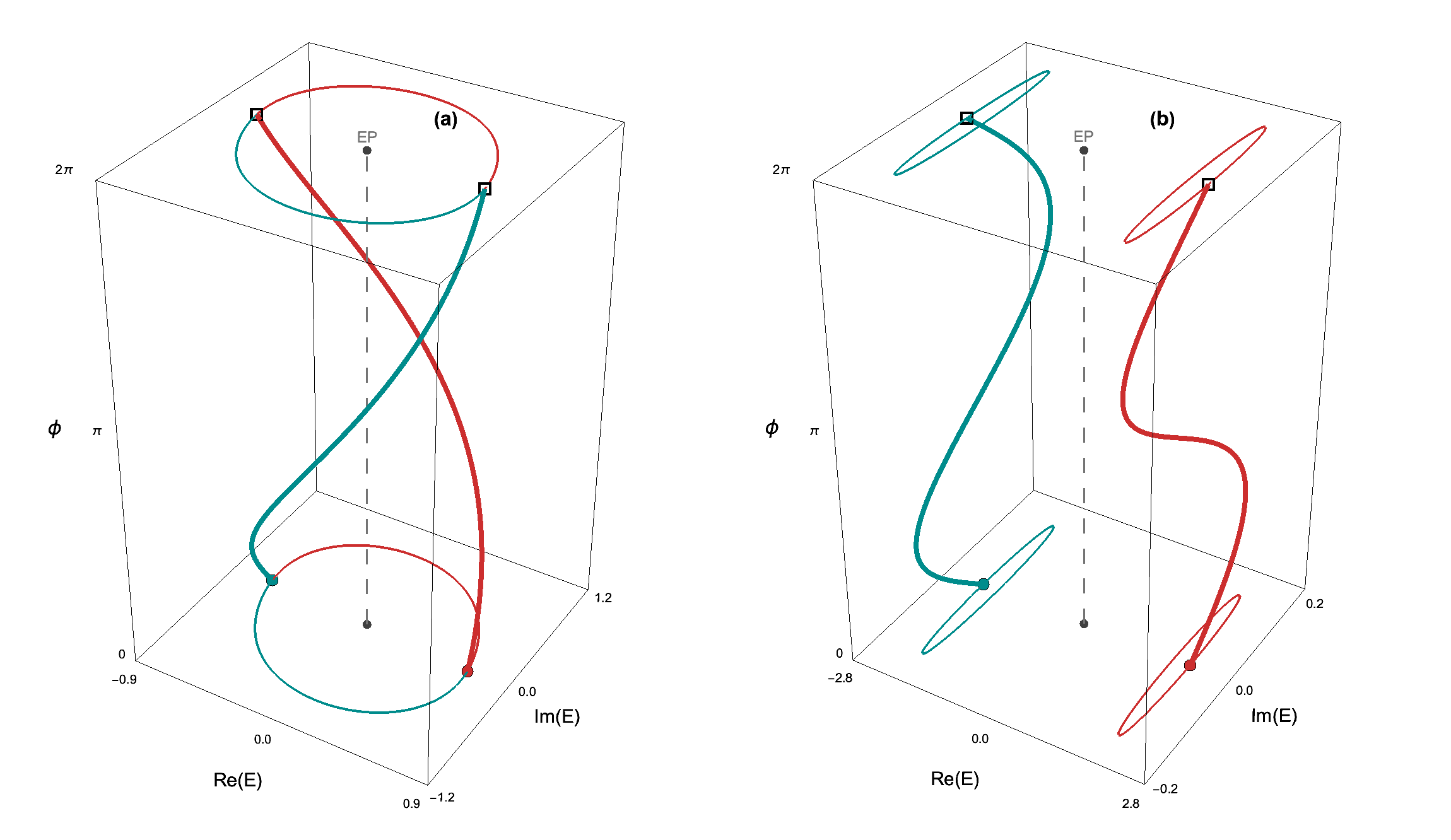}
\caption{Eigenvalue braiding associated with the switching domains shown in Fig.~\ref{fig:domains}(a). 
(a) Momentum point inside the switching domain ($W=1$). The eigenvalues are shown in the three-dimensional space $(\mathrm{Re}E,\mathrm{Im}E,\phi)$ during one modulation cycle $\phi:0\rightarrow2\pi$. The eigenvalue trajectories braid around the EP, leading to an exchange of eigenmodes after one cycle. 
(b) Momentum point outside the switching domain ($W=0$). The eigenvalues evolve without braiding and return to their original branches after the cycle. 
Filled circles indicate the initial eigenvalues at $\phi=0$, and open squares indicate the final eigenvalues at $\phi=2\pi$.}
\label{fig:braiding}
\end{figure}

Figure~\ref{fig:braiding} illustrates the eigenvalue braiding associated with the switching domains during one modulation cycle of the control parameter $\phi$. For a momentum point inside the switching domain ($W=1$), the eigenvalues braid around the EP during the cycle and exchange after one period. In contrast, for a momentum point outside the switching domain ($W=0$), the eigenvalues evolve without braiding and return to their original branches. These results confirm that the invariant $W(\mathbf{k})$ captures the momentum-dependent band permutation induced by cyclic modulation [SM-E].

\begin{figure*}
\centering
\includegraphics[width=1.\textwidth]{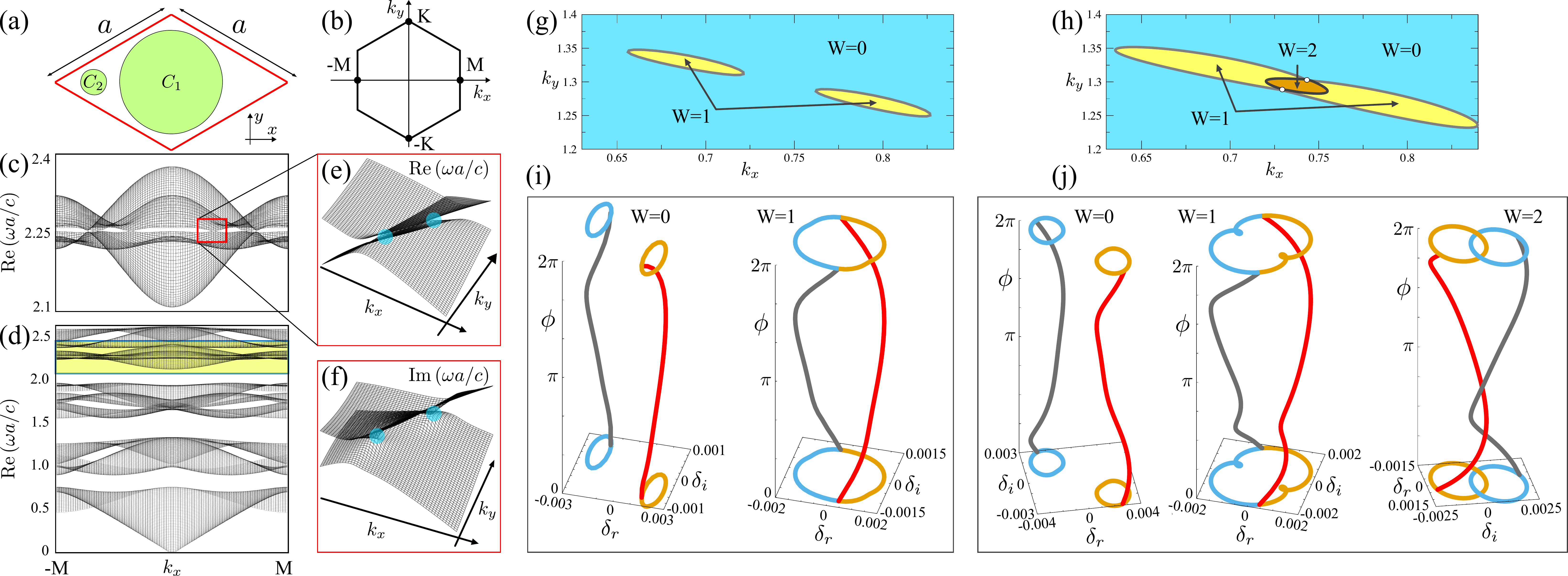}
\caption{Photonic realization of momentum-space switching domains. (a) Schematic of the photonic-crystal unit cell consisting of a rhombic primitive cell of a triangular lattice with two dielectric disks $C_1$ and $C_2$. (b) Corresponding Brillouin zone with high-symmetry points $[\mathbf{M},\mathbf{K}]$. (c) Real part of the eigenfrequency for the coupled two-level band structure, extracted by isolating the band manifold within the shaded-box region in (d), which hosts an EP pair. (d) The projected real part of the eigenfrequency for the bands computed over the two-dimensional Brillouin zone, in the range Re$(\omega a/c)\in[0,2.65]$.
(e,f) Enlarged views of the real and imaginary parts of the eigenfrequency near the EP pair. The EPs are marked by solid circles. (g,h) Momentum-space distributions of the winding number $W(\mathbf{k})$ for representative modulation amplitudes $\mu$, showing regions with $W=0,1,2$. The domain boundaries coincide with the projections of the EP trajectories. Here (g) and (h) correspond to $\mu=0.125$ and $\mu=0.25$, respectively. (i,j) Representative trajectories of the complex eigenfrequencies as a function of $\phi$ for selected momentum points corresponding to the regions in (g) and (h). The frequencies are shown relative to the band average, $\delta^\pm=\omega^j a/c-\langle \omega a/c\rangle$, with a band index $j=1,2$.}
\end{figure*}

\textcolor{blue}{\textit{Photonic implementation -}}
The switching-domain physics developed above is general and does not rely on specific symmetries, models, or parameter choices, arising from the presence of EPs and a cyclically modulated external parameter. As a representative implementation, we consider a photonic crystal with lossy dielectric materials, which naturally realizes non-Hermitian band structures with momentum-space EPs \cite{Cerjan2016exceptional, Chen2019tailoring, Zhong2023numerical, Ryu2025pseudo}.

We study a two-dimensional photonic crystal based on a triangular lattice with a rhombic primitive cell containing two dielectric disks of radii $R_{C_1}=2a/3\sqrt{3}$ and $R_{C_2}=a/6\sqrt{3}$, located at $(0,0)$ and $(-a/\sqrt{3},0)$, respectively [Fig.~4(a)]. To obtain the complex-valued photonic bands, we solve the two-dimensional Helmholtz equation,
\begin{equation}
\nabla^2 \Psi(\mathbf{r}) + \frac{\omega^2}{c^2} n^2(\mathbf{r}) \Psi(\mathbf{r}) = 0,
\label{eq:numeric}
\end{equation}
with Bloch periodic boundary conditions. To numerically obtain the eigenfrequencies $(\omega\in\mathbb C)$ and corresponding eigenmodes $(\Psi\in\mathbb C)$ that satisfy Eq.~(\ref{eq:numeric}), we implement the boundary element method~\cite{Jan_Wiersig_2003,LI2012527}. The detailed procedures are the same as the ones found in \cite{BEM_bands1,BEM_bands2}. 
Here, we examined the system with the refractive indices given as $n^0_{C_1}=3.097+i0.930$ and $n^0_{C_2}=1.597+i0.440$. These parameters are chosen such that a pair of EPs emerges within a selected two-band manifold [Figs.~4(c)–(f)]. The positive imaginary values in the refractive index imply the absorption loss. The cyclic control parameter $\phi$ is introduced by varying the complex refractive index of disk $C_1$ as $n_{C_1}(\phi)=n^0_{C_1}+\mu(1-e^{i\phi})$, thereby tracing a closed trajectory in the complex plane. Here, $\mu$ controls the size of the trajectory in the complex parameter space and determines whether the EP loops remain separated or overlap.

We compute the complex-valued band structure over the full two-dimensional Brillouin zone. For visualization, the real part of the eigenfrequency is projected onto the $k_x$ axis by aggregating over $k_y$ [Figs.~4(c),(d)], while the local band structure near the EP pair is shown in detail in Figs.~4(e),(f). The band-permutation invariant $W(\mathbf{k})$ is defined from the phase winding of the squared band gap $(\delta^+ - \delta^-)^2$ and directly determines whether eigenmodes are permuted after one modulation cycle. As the control parameter $\phi$ evolves, the EPs move in the extended parameter space $(k_x,k_y,\phi)$, and their projections onto the Brillouin zone coincide with the boundaries between regions with different values of the band-permutation invariant $W(\mathbf{k})$ [Figs.~4(g),(h)]. Note that the EP trajectories on $(k_x,k_y)$-plane are computed following the control parameter $\phi$ with a numerical tolerance $\delta_\text{EP}(k_x,k_y)=|\omega^1(k_x,k_y)-\omega^2(k_x,k_y)|<10^{-9}$.

Figures~4(g) and 4(h) show the resulting switching-domain maps for two representative modulation amplitudes. For the smaller modulation amplitude ($\mu=0.125$), the projected EP trajectories form separated narrow domains, so that the Brillouin zone is divided mainly into nonswitching regions with $W=0$ and localized switching regions with $W=1$ [Fig.~4(g)]. For the larger modulation amplitude ($\mu=0.25$), the projected trajectories expand and overlap, producing a richer domain pattern that includes regions with $W=2$ [Fig.~4(h)]. Thus, in the photonic-crystal realization, increasing the modulation amplitude reorganizes the momentum-space topology in the same way as in the minimal model, but with additional higher-winding domains arising from the more complicated band structure.

The representative complex-frequency trajectories in Figs.~4(i) and 4(j) directly illustrate the physical meaning of these domains. In the $W=0$ region, the two eigenfrequency branches return to themselves after one modulation cycle (i.e., $\phi:0\to2\pi$), indicating trivial band permutation. In the $W=1$ region, the branches are exchanged after the cycle, demonstrating eigenmode switching. In the $W=2$ region, the trajectories wind twice in terms of the squared band gap, corresponding to a full winding of the eigenvalues without net band permutation. Although the braiding is nontrivial during the cycle, the branches return to their original labels after one period. These results confirm that the integer $W(\mathbf{k})$ correctly predicts the band permutation and braiding behavior in the photonic system.

\textcolor{blue}{\textit{Conclusion -}}
We have shown that mobile EPs generate momentum-space switching domains in non-Hermitian band systems. By introducing a band-permutation invariant, we demonstrated that cyclic modulation partitions the Brillouin zone into regions with distinct eigenmode-switching behavior. The boundaries of these domains are determined by the projections of EP trajectories in the extended parameter space $(k_x,k_y,\phi)$, establishing a direct connection between EP motion and band permutation topology.

As the modulation strength increases, the switching domains evolve from isolated regions to extended structures that can cover the Brillouin zone, leading to global band switching. In more general band structures, overlapping EP trajectories can also give rise to higher winding numbers, where nontrivial eigenvalue braiding may occur even when the net band permutation is trivial. We further demonstrated this mechanism in a photonic crystal with lossy dielectric material, showing that the proposed switching-domain physics extends beyond minimal models and can be realized in realistic wave systems. These results provide a route for engineering non-Hermitian spectral topology through controllable EP motion.

\acknowledgments
We acknowledge financial support from the Institute for Basic Science in the Republic of Korea through the projects IBS-R041-A2-2026-a00 and IBS-R041-D1-2026-a00. C.-H. Yi acknowledges a financial support from the National Research Foundation of Korea (NRF) grant funded by the Korea government (MSIT) (Grants No. RS-2025-16070482, RS-2025-25464760, RS-2023-NR119928, RS-2025-25446099, RS-2025-03392969, RS-2023-00278511, RS-2025-02315685)

\bibliography{reference}

\clearpage
\onecolumngrid
\newpage

\setcounter{figure}{0}
\setcounter{table}{0}
\setcounter{equation}{0}
\setcounter{subsection}{0}

\renewcommand{\thefigure}{S\arabic{figure}}
\renewcommand{\thetable}{S\arabic{table}}
\renewcommand{\theequation}{S\arabic{equation}}
\renewcommand{\thesubsection}{\Alph{subsection}}

\section*{Supplemental Material}

\subsection{A. Analytic condition and geometric interpretation of the switching boundary}

The boundary separating switching and nonswitching regions can be obtained analytically from Eqs.~(1) and (2) of the main text. 
EPs occur when the two eigenvalues coalesce, corresponding to the condition
\begin{equation}
\Delta(\mathbf{k},\phi)=0 .
\end{equation}

Using Eq.~(2) of the main text, this condition becomes
\begin{equation}
\epsilon_0^2+(t_x+t_y)
\left(
t_x e^{ik_x}+t_y e^{ik_y}+\mu e^{i\phi}
\right)=0 .
\end{equation}

Rearranging terms gives
\begin{equation}
\epsilon_0^2+(t_x+t_y)(t_x e^{ik_x}+t_y e^{ik_y})
+(t_x+t_y)\mu e^{i\phi}=0 .
\end{equation}

Defining
\begin{equation}
C(\mathbf{k})=
\epsilon_0^2+(t_x+t_y)(t_x e^{ik_x}+t_y e^{ik_y}),
\end{equation}

the EP condition can be written as
\begin{equation}
C(\mathbf{k})+(t_x+t_y)\mu e^{i\phi}=0 .
\end{equation}

Since $|e^{i\phi}|=1$, a solution for $\phi$ exists only when
\begin{equation}
|C(\mathbf{k})|=|t_x+t_y|\mu .
\end{equation}

This condition determines the set of momenta for which the spectral trajectory $\Delta(\mathbf{k},\phi)$ reaches the origin of the complex plane for some value of $\phi$. 
At these momenta the spectral gap closes and the winding number $W(\mathbf{k})$ can change, thereby defining the boundary between switching and nonswitching regions in the Brillouin zone.

Consequently, the switching boundary corresponds to the projection of EP trajectories in the extended parameter space $(k_x,k_y,\phi)$ onto the $(k_x,k_y)$ plane. 
Across this boundary, the winding number $W(\mathbf{k})$ changes between 0 and 1.

This result can be further understood by explicitly evaluating the winding number. 
From Eq.~(S5), the spectral quantity can be written as
\begin{equation}
\Delta^2(\mathbf{k},\phi)
=
C(\mathbf{k})+(t_x+t_y)\mu e^{i\phi},
\end{equation}
which describes a circular trajectory in the complex plane centered at $C(\mathbf{k})$ with radius $|(t_x+t_y)\mu|$.

When $|C(\mathbf{k})|<|(t_x+t_y)\mu|$, the trajectory of $\Delta^2(\mathbf{k},\phi)$ encloses the origin during one modulation cycle, yielding $W(\mathbf{k})=1$.

In contrast, when $|C(\mathbf{k})|>|(t_x+t_y)\mu|$, the trajectory does not enclose the origin, yielding $W(\mathbf{k})=0$.

Therefore, the condition $|C(\mathbf{k})|=|(t_x+t_y)\mu|$ determines the switching boundary, separating the regions with $W(\mathbf{k})=1$ and $W(\mathbf{k})=0$.

\subsection{B. Numerical evaluation of the band-permutation invariant}

In the numerical calculations presented in the main text, the band-permutation invariant $W(\mathbf{k})$ is obtained from the winding of $\Delta^2(\mathbf{k},\phi)$ during one modulation cycle. 
To ensure a well-defined invariant, we evaluate the winding using the squared quantity $\Delta^2(\mathbf{k},\phi)$, which is single-valued over the cycle. 
For each momentum point $\mathbf{k}$, the control parameter $\phi$ is discretized over the interval $[0,2\pi]$, and the complex quantity $\Delta^2(\mathbf{k},\phi)$ is evaluated along the cycle.

The invariant is computed from the phase winding of $\Delta^2(\mathbf{k},\phi)$ around the origin of the complex plane,
\begin{equation}
W(\mathbf{k})
=
\frac{1}{2\pi}
\oint d\phi \,
\partial_\phi
\arg\!\left[\Delta^2(\mathbf{k},\phi)\right].
\end{equation}

In practice, this winding number is evaluated numerically using a discretized phase accumulation,
\begin{equation}
W(\mathbf{k})
=
\frac{1}{2\pi}
\sum_j
\arg\!\left[
\frac{\Delta^2(\mathbf{k},\phi_{j+1})}
{\Delta^2(\mathbf{k},\phi_j)}
\right],
\end{equation}
where the phase difference is obtained from the argument of the complex ratio.

A winding number $W(\mathbf{k})=1$ indicates that $\Delta^2(\mathbf{k},\phi)$ encloses the origin in the complex plane, corresponding to a sign change of $\Delta(\mathbf{k},\phi)$ after one modulation cycle and hence eigenmode exchange. 
In contrast, $W(\mathbf{k})=0$ indicates that the trajectory does not enclose the origin and therefore no band permutation occurs. 
This numerical procedure is used to construct the momentum-space maps of $W(\mathbf{k})$ presented in Fig.~2 of the main text. 
The discretization of the modulation cycle was chosen sufficiently fine to ensure convergence of the winding number.

\subsection{C. EP trajectory in extended parameter space}

\begin{figure*}
\centering
\includegraphics[width=1.0\linewidth]{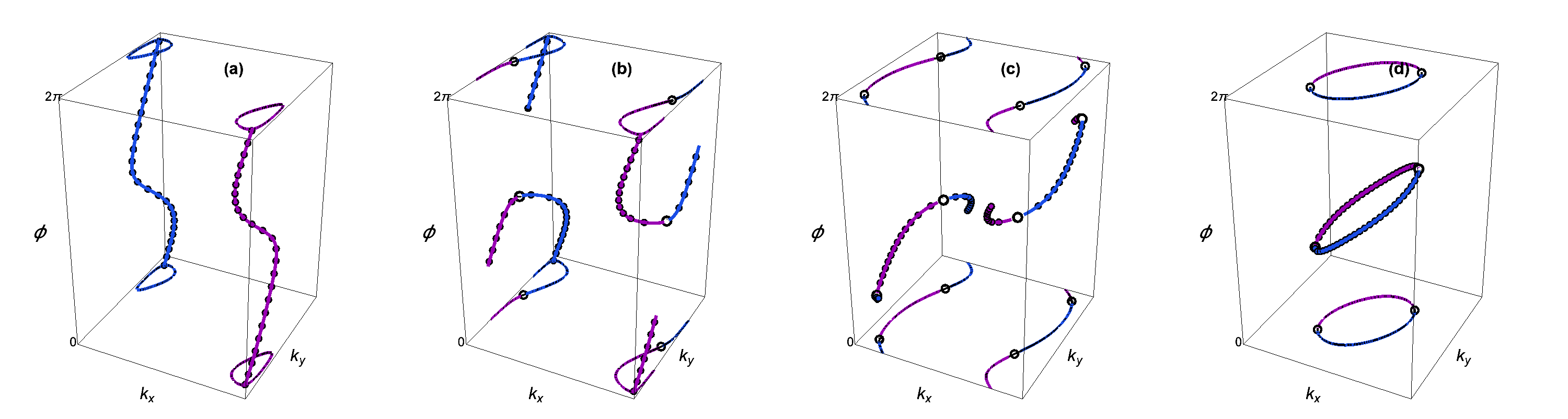}
\caption{Trajectories of EPs in the extended parameter space $(k_x,k_y,\phi)$. Blue and purple curves denote the two EP branches as the cyclic control parameter $\phi$ varies from $0$ to $2\pi$. Open circles indicate pair-annihilation and pair-creation points where the two EPs merge and disappear or re-emerge. Panels (a-d) correspond to the same parameter regimes as in Fig.~2 of the main text. The projections of these trajectories onto the $(k_x,k_y)$ plane coincide with the switching-domain boundaries of the band-permutation invariant $W(\mathbf{k})$.}
\label{fig:figS1}
\end{figure*}

To visualize the origin of the switching domains, we analyze the motion of EPs in the extended parameter space $(k_x,k_y,\phi)$. The EP condition yields two solution branches that trace continuous trajectories as $\phi$ evolves from $0$ to $2\pi$. Figure~\ref{fig:figS1} shows the resulting EP trajectories for the same parameter regimes as those used in Fig.~2 of the main text. As $\phi$ varies, the EPs move through the Brillouin zone and form curves in the extended parameter space. Open circles mark collision points where the two EPs merge and annihilate or reappear through pair creation. 

\subsection{D. EP regimes of the model}

\begin{figure*}[t]
\centering
\includegraphics[width=0.6\linewidth]{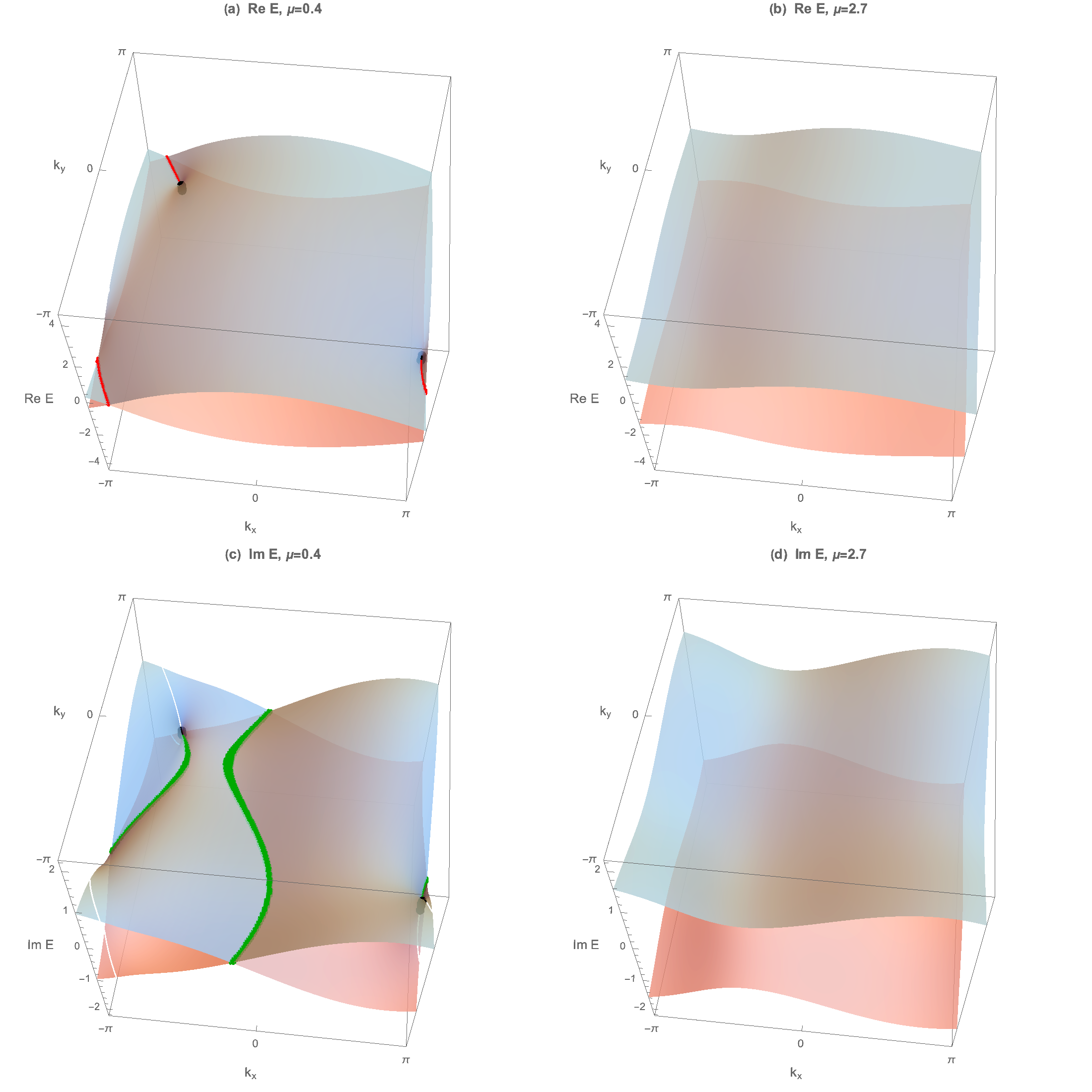}
\caption{
Limiting spectral regimes for small and large $\mu$, corresponding to the presence and absence of EPs.
(a,c) For $\mu=0.4$, two EPs are present in the Brillouin zone and the two sheets of the spectrum are connected by branch cuts terminating at the EPs.
(b,d) For $\mu=2.7$, the spectrum remains fully gapped and no EPs occur, so the two sheets remain disconnected.
Red (green) points indicate the loci where the real (imaginary) part of the band gap vanishes, corresponding to the branch cuts of the real (imaginary) Riemann surfaces, while black points denote EPs.
All panels are calculated for $t_x=1$, $t_y=0.5$, and $\epsilon_0=1.2$ with $\phi=\pi/2$.
}
\end{figure*}

\begin{figure*}[t]
\centering
\includegraphics[width=0.95\linewidth]{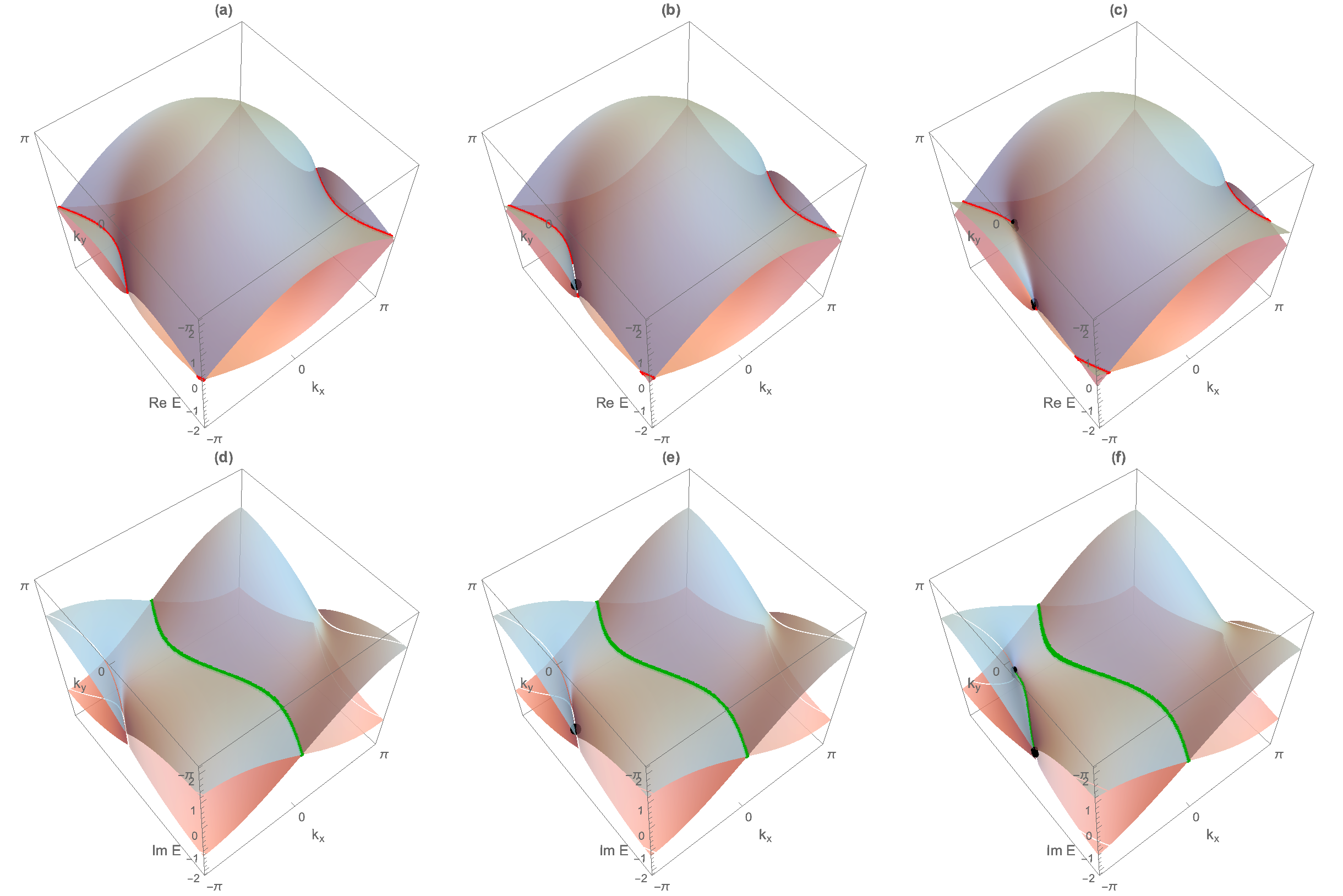}
\caption{
EP annihilation and recreation near the transition for $\mu=0.5$.
(a,d) At $\phi=3.0$, the spectrum is fully gapped and no EPs are present.
(b,e) At the critical value $\phi=2.8578$, the gap closes at a single momentum point.
(c,f) At $\phi=2.4$, a pair of EPs is present and the two sheets of the spectrum are connected by branch cuts terminating at the EPs.
Red (green) points indicate the loci where the real (imaginary) part of the band gap vanishes.
All panels use the parameters $t_x=1$, $t_y=0.5$, $\epsilon_0=1.2$, and $\mu=0.5$.
}
\end{figure*}

To illustrate the different spectral regimes of the model, we examine representative Riemann surfaces of the complex energy spectrum. In the numerical calculations shown in Supplementary Figs.~S2 and S3 we use the representative parameters $t_x=1$, $t_y=0.5$, and $\epsilon_0=1.2$. Supplementary Fig.~S2 shows two limiting cases of the modulation amplitude $\mu$. For sufficiently small $\mu$, two EPs are present in the Brillouin zone for all values of the cyclic parameter $\phi$. In this regime the two sheets of the spectrum are connected through branch cuts terminating at the EPs. In contrast, for sufficiently large $\mu$ the spectrum remains fully gapped and no EPs exist in the Brillouin zone. In this case the two sheets of the spectrum remain disconnected.

EPs occur when the eigenvalue discriminant of the Hamiltonian vanishes, i.e., when the two complex eigenvalues become degenerate. For the present model this condition can be written as
\[
|t_x-t_y|
\le
\left|
\frac{\epsilon_0^2}{t_x+t_y}+\mu e^{i\phi}
\right|
\le
t_x+t_y .
\]

For the parameters used in this work this condition becomes
\[
0.5 \le |0.96+\mu e^{i\phi}| \le 1.5 .
\]

This inequality defines an annular region in the complex plane of the parameter $0.96+\mu e^{i\phi}$. EPs exist only when the circular trajectory of this parameter intersects this annulus as $\phi$ varies from $0$ to $2\pi$.

For the present parameters the structure of EPs changes at two characteristic modulation amplitudes. When $\mu<0.46$, the circular trajectory remains entirely inside the annulus and two EPs exist for all values of $\phi$. When $\mu>2.46$, the trajectory lies completely outside the annulus and the spectrum remains fully gapped throughout the entire modulation cycle. In the intermediate regime $0.46<\mu<2.46$, the trajectory intersects the annulus only within finite intervals of $\phi$, leading to the creation and annihilation of EP pairs during the cycle.

For representative values of the modulation amplitude used in Fig.~2, Table~\ref{tab:EP_phi_structure} summarizes the presence or absence of EPs during the modulation cycle $\phi:0\rightarrow2\pi$. These intervals can be directly compared with the EP trajectories shown in Supplementary Fig.~S1.

\begin{table}[t]
\caption{EP structure during the modulation cycle $\phi:0\rightarrow2\pi$
for representative values of $\mu$ used in Fig.~2.}
\begin{ruledtabular}
\begin{tabular}{c c}
$\mu$ & EP structure during the cycle \\
\hline
0.4 & EPs present for all $\phi$ \\

0.5 & EPs present for $0\le\phi<2.86$ and $3.43<\phi\le2\pi$; \\
    & no EPs for $2.86<\phi<3.43$ \\

0.9 & EPs present for $1.27<\phi<2.60$ and $3.68<\phi<5.02$; \\
    & no EPs otherwise \\

2.0 & EPs present for $2.34<\phi<3.94$; \\
    & no EPs otherwise \\
\end{tabular}
\end{ruledtabular}
\label{tab:EP_phi_structure}
\end{table}

The branch cuts visible in the spectral surfaces originate from the multi-valued nature of the complex eigenvalues near EPs. They represent the loci in momentum space where the real or imaginary part of the band gap vanishes, thereby connecting the two sheets of the Riemann surface.

As $\phi$ varies during the modulation cycle, the EPs move through the Brillouin zone and may annihilate or reappear depending on the value of $\mu$. Supplementary Fig.~S3 illustrates this process for $\mu=0.5$. In the panels shown there, the spectrum is fully gapped at $\phi=3.0$, the gap closes at the critical value $\phi=2.8578$, and two EPs are present at $\phi=2.4$. This evolution reflects the annihilation or recreation of an EP pair as the cyclic parameter is varied.

These results demonstrate that the switching-domain structure discussed in the main text originates from the geometric motion of EPs in the extended parameter space $(k_x,k_y,\phi)$. As the control parameter $\phi$ evolves during the cycle, the EPs move through momentum space and reorganize the spectral Riemann surface.

\subsection{E. Additional examples of eigenvalue braiding} 

\begin{figure}
\centering
\includegraphics[width=0.6\linewidth]{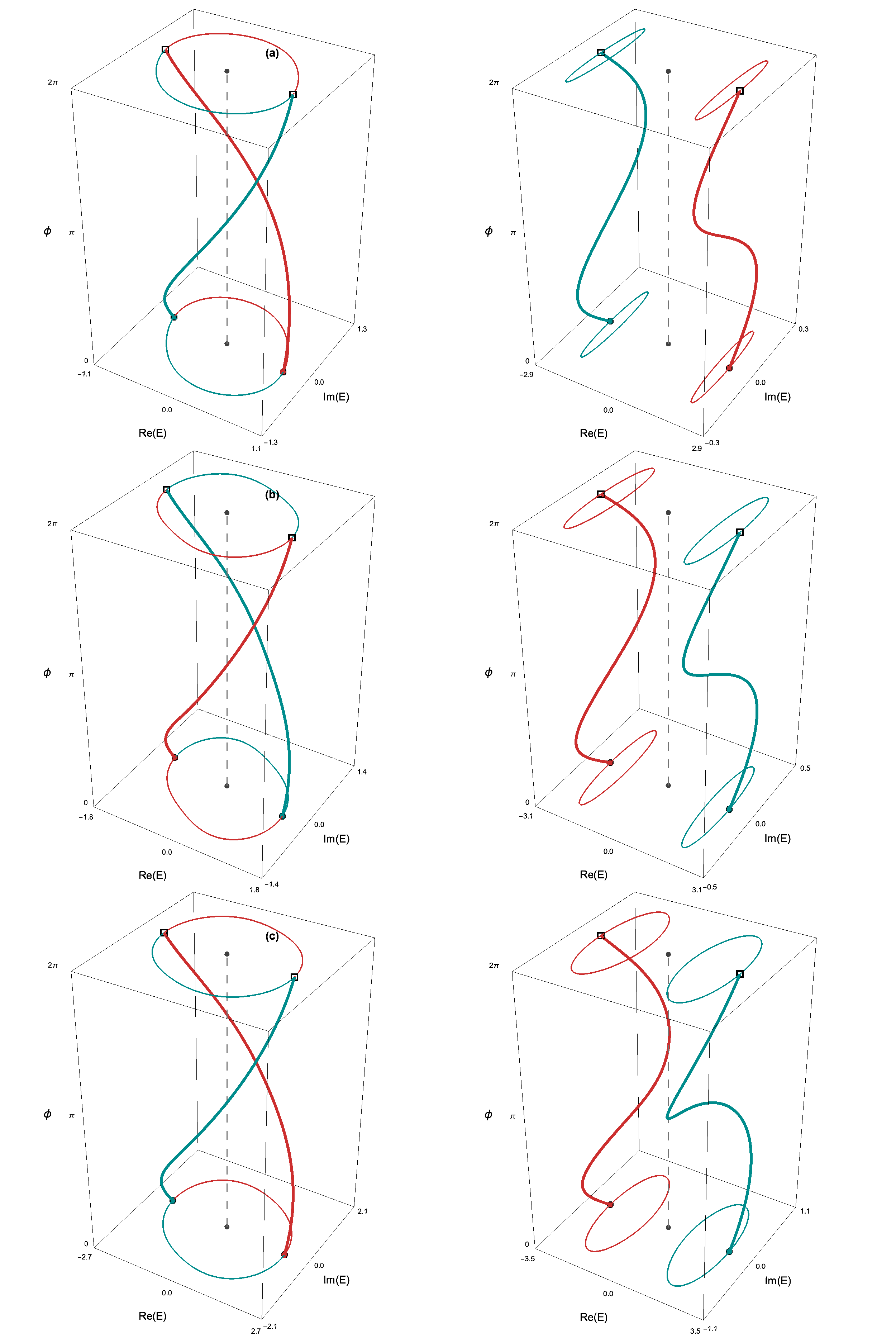}
\caption{
Complex-energy braiding for representative momentum points corresponding to the domain configurations in Fig.~2(b)–(d). 
Panels (a), (b), and (c) correspond to representative momentum points from the domain configurations shown in Fig.~2(b), Fig.~2(c), and Fig.~2(d), respectively. 
For each value of $\mu$, the left panel shows a momentum point with $W(\mathbf{k})=1$, while the right panel shows a point with $W(\mathbf{k})=0$. 
The solid curves represent the eigenvalue trajectories in the three-dimensional space $(\mathrm{Re}E,\mathrm{Im}E,\phi)$ during one modulation cycle $\phi:0\rightarrow2\pi$. 
Filled circles indicate the eigenvalues at $\phi=0$, and open squares indicate the eigenvalues at $\phi=2\pi$. 
For $W(\mathbf{k})=1$, the trajectories encircle the origin, resulting in an exchange of eigenmodes after the cycle, whereas for $W(\mathbf{k})=0$, the trajectories do not enclose the origin and the eigenvalues return to their original branches.
}
\label{fig:figS4}
\end{figure}

In the main text we demonstrated the relation between the band-permutation invariant and eigenvalue braiding using representative momentum points corresponding to the domain structure in Fig.~2(a). 
Here we provide additional examples corresponding to the domain configurations shown in Fig.~2(b)–(d).

Figure~\ref{fig:figS4} shows the complex-energy trajectories for representative momentum points with $W(\mathbf{k})=1$ and $W(\mathbf{k})=0$. 
These examples further confirm that the invariant $W(\mathbf{k})$ determines whether eigenvalue braiding occurs during cyclic modulation. 
The eigenvalue branches are tracked continuously along the modulation cycle by matching eigenvalues between neighboring $\phi$ steps in the complex-energy plane.

\subsection{F. Supplementary movies}
    
To provide a direct visualization of the spectral evolution discussed in the main text, we provide two sets of animations illustrating the motion of EPs and the resulting evolution of the spectral Riemann surfaces during one modulation cycle of the control parameter $\phi$.
    
These movies provide a direct visualization of the geometric origin of the switching domains, where the boundaries correspond to the projection of EP trajectories in the extended parameter space $(k_x,k_y,\phi)$ onto the $(k_x,k_y)$ plane.

\textbf{Movies S1–S5: EP motion and branch-cut evolution in the Brillouin zone.}
    
These animations show the motion of EPs and the corresponding branch cuts in the Brillouin zone as the control parameter $\phi$ varies from $0$ to $2\pi$. 
The black curves denote the boundaries of the switching domains determined by the band-permutation invariant $W(\mathbf{k})$. 
Red and green curves indicate the real and imaginary branch cuts, respectively, and black points denote EPs. 
The red and blue regions correspond to momentum-space domains where band permutation occurs ($W=1$) and does not occur ($W=0$). 
These animations illustrate how the motion of EPs and the accompanying evolution of branch cuts in the Brillouin zone generate the switching-domain structure in momentum space.
    
\textit{Movie S1:} $\mu=0.4$ \; (\texttt{mu0.4\_animation.gif})  
    
\textit{Movie S2:} $\mu=0.5$ \; (\texttt{mu0.5\_animation.gif})  
  
\textit{Movie S3:} $\mu=0.9$ \; (\texttt{mu0.9\_animation.gif})  
    
\textit{Movie S4:} $\mu=2.0$ \; (\texttt{mu2.0\_animation.gif})  
    
\textit{Movie S5:} $\mu=2.7$ \; (\texttt{mu2.7\_animation.gif})

\textbf{Movies S6–S10: Evolution of the complex band structure.}
    
These animations show the real and imaginary parts of the complex bands during one modulation cycle of $\phi$ for representative values of the modulation amplitude $\mu$. 
The two bands are displayed as surfaces over the Brillouin zone together with the positions of EPs and the corresponding branch cuts, illustrating how the Riemann-surface structure evolves as $\phi$ varies.
    
In each animation two representative momentum points are tracked. 
The orange marker denotes a point located inside the band-permutation domain, while the blue marker denotes a point outside this domain. 
As $\phi$ evolves over one cycle, the tracked eigenmode at the orange point moves from one band to the other, demonstrating band permutation. 
In contrast, the eigenmode at the blue point returns to the same band after the cycle.
    
\textit{Movie S6:} $\mu=0.4$ \; (\texttt{riemann\_real\_imag\_animation\_mu0.4.gif})
    
\textit{Movie S7:} $\mu=0.5$ \; (\texttt{riemann\_real\_imag\_animation\_mu0.5.gif})
    
\textit{Movie S8:} $\mu=0.9$ \; (\texttt{riemann\_real\_imag\_animation\_mu0.9.gif})
    
\textit{Movie S9:} $\mu=2.0$ \; (\texttt{riemann\_real\_imag\_animation\_mu2.0.gif})
    
\textit{Movie S10:} $\mu=2.7$ \; (\texttt{riemann\_real\_imag\_animation\_mu2.7.gif})


\end{document}